# Superconducting MAX phases $Nb_2AC$ (A = S, Sn): An *ab-initio* study

## M. T. Nasir, A.K.M.A. Islam[*]

*Department of Physics, Rajshahi University, Rajshahi 6205, Bangladesh*

**Abstract**

An *ab-initio* investigation of structural parameters, elastic, electronic, thermodynamic and optical properties of MAX phases $Nb_2AC$ (A = S, Sn) has been carried out by the plane wave psedudopotential method based on density functional theory (DFT). The effect on results of substitution of heavier Sn atoms for the lighter S atoms in the nanolaminate network has been made. The analysis of the electronic band structure shows that these compounds are electrical conductors, with contribution predominantly from the Nb $4d$ states. The temperature and pressure dependence of bulk modulus, Debye temperature, specific heats, thermal expansion coefficient of the nanolaminates are calculated for the first time using the quasi-harmonic Debye model with phononic effects. The estimated values of electron-phonon coupling constants ( $\lambda \sim 0.49$, $\sim 0.59$) imply that $Nb_2SC$ and $Nb_2SnC$ are moderately coupled superconductors. Further first time detailed analysis of all optical functions reveals that $Nb_2SC$ is a better dielectric material, and also both the phases, having similar reflectivity spectra, show promise as good coating materials in the energy regions 10- 16.5 eV.

*Keywords*: $Nb_2SC$; $Nb_2SnC$; First-principles; Mechanical properties; Quasi-harmonic Debye model; Band structure; Thermodynamic properties; Optical properties

## 1. Introduction

The group of materials known as $M_{n+1}AX_n$ (MAX) phases since after the discovery by Nowotny *et al.* [1] have attracted great interest in the scientific communities worldwide [2-26] owing to a remarkable combination of properties, which are characteristic both of metals and ceramics. Furthermore out of a total of about 60 synthesized MAX phases [3], seven low-$T_c$ superconductors have so far been identified. These are: $Mo_2GaC$ [4], $Nb_2SC$ [5], $Nb_2SnC$ [6], $Nb_2AsC$ [7], $Ti_2InC$ [8], $Nb_2InC$ [9], and $Ti_2InN$ [10].

Sakamaki *et al.* [5] in 1999 reported the discovery of a new class of layered superconductor known as carbosulfide, $Nb_2SC$, which was in fact synthesized in 1968 by Beckmann *et al.* [11]. The electronegativity of C is same as S and as a result C-containing sulfide will be of particular interest in that it may cause a change in the structural dimensionality and physical properties of sulfide. Bortolozo *et al.* [6] report a study of the conductivity properties of $Nb_2SnC$ sintered at both ambient and high pressure. The investigation revealed that this compound shows superconducting transition at

---
[*] Corresponding author. Tel.: +88 0721 750980; fax: +88 0721 750064.
   *E-mail address*: azi46@ru.ac.bd (A.K.M.A. Islam).

~ 7.8 K. The sample, when heat treated under high pressure, exhibited sharp $T_c$ in both transport and magnetization properties. Prior to this El-Raghy *et al.* [26] synthesized $Nb_2SnC$ along with other MAX phases and made several measurements.

The two MAX phases under consideration in this study possess a layered hexagonal structure (where blocks of transition metal carbides [MC] formed by edge-shared $M_6C$ octahedra are sandwiched with A atomic sheets) and thus are similar to other groups of layered SCs such as high-$T_C$ oxides, $MgB_2$. Halilov *et al.* [12] suggested that superconductivity in $Nb_2SC$ arises due to pairing of electrons of S atoms arranged in planar networks. Since, according to the BCS theory $T_c$ ~(atomic mass)$^{-1/2}$, the higher $T_c$ of $Nb_2SnC$ as compared to that of $Nb_2SC$ (substitution of heavier Sn for the lighter S) clearly indicates that the pairing mechanism in these phases should have a different nature. Shein *et al.* [13] believe that the major role in the superconductivity of nanolaminates is played by the states of carbide (Nb–C) molecular layers. These layers, like the structure of the known superconductor NbC ($T_c$~ 11 K) are composed of [$Nb_6C$] octahedra.

The electronic structures of $Nb_2SC$ and $Nb_2SnC$ have been carried out by Shein *et al.* [13] using FLAPW within GGA. Also Bouhemadou [14] has performed first-principles calculations for the structural, elastic properties of only $Nb_2SnC$ under pressure by employing PP-PW approach based on DFT within the local density approximation (LDA). Kanoun *et al.* [17] investigated structural, elastic, electronic and dielectric function of four 211 MAX phases including $Nb_2SnC$. Shein *et al.* [15] presented a theoretical study of the elastic properties for six superconducting MAX phases: $Nb_2SC$, $Nb_2SnC$, $Nb_2AsC$, $Nb_2InC$, $Mo_2GaC$, and $Ti_2InC$ using VASP code in projector augmented wave formalism. Recently Romero *et al.* [16] studied the pressure dependence of elastic and electronic properties of $Nb_2SnC$.

In view of the above discussions, it is clear that $Nb_2SC$ has been subjected to limited theoretical study. Furthermore full optical as well as finite-temperature and finite-pressure thermodynamical investigations are absent for both the nanolaminates. The situation demands focus on areas where little or no work has been carried out. All these motivate us to perform such calculations on $Nb_2SC$ and $Nb_2SnC$ in addition to revisiting the existing theoretical works using methodology different from those used in previous published works [13-15]. The parameters of optical properties (dielectric function, absorption spectrum, conductivity, energy-loss spectrum, refractive index and reflectivity) for both the phases will be calculated and discussed. The paper is divided in four sections. In Sec. 2, we briefly describe the computational techniques used in this study. The results obtained for the structural, elastic, electronic, thermodynamic and optical properties for $Nb_2SC$ and $Nb_2SnC$ phases are presented and discussed in Sec. 3. Finally, Sec. 4 summarizes the main conclusion of the present work.

## 2. Computational methods

The zero temperature energy calculations presented in this work were performed by employing CASTEP code [27] which utilizes the plane-wave pseudopotential based on the framework of density functional theory (DFT). The electronic exchange-correlation energy is treated under the generalized gradient approximation (GGA) in the scheme of Perdew-Burke-Ernzerhof (PBE) [28]. The interactions between ion and electron are represented by ultrasoft Vanderbilt-type pseudopotentials for Nb, S, Sn and C atoms [29]. The basis set of valence electronic states was set to be $4d^45s^1$ for Nb, $3s^23p^4$ for S, $4d^{10}5s^25p^2$ for Sn, $2s^22p^2$ for C. The correlation functional GGA-PBE was used in the calculations plane-wave basis set with 500 eV energy cut-off. For the sampling of the Brillouin zone a Monkhorst-Pack grid [30] of 10×10×3 k-points for Nb$_2$SC, 13×13×3 k-points for Nb$_2$SnC were employed. All the structures were relaxed by BFGS methods [31]. For the geometry optimization, the convergence tolerances were set as follows: 1×10$^{-5}$ eV/atom for the total energy, 0.002 eV/Å for the maximum force on atoms, 0.05 GPa for the maximum stress, 1×10$^{-3}$ Å for the maximum atomic displacement.

The thermodynamic properties of Nb$_2$SC and Nb$_2$SnC were investigated using the quasi-harmonic Debye model, the detailed description of which can be found elsewhere [32, 33]. Here we have calculated the bulk modulus, volume thermal expansion coefficient (VTEC), specific heats and Debye temperature at different temperatures and pressures. For this purpose we used $T = 0$K, $P = 0$ GPa energy-volume calculation based on DFT method to obtain necessary $E$-$V$ data employing the third order Birch-Murnaghan EOS [34].

## 3. Results and discussion

### 3.1. Structural and elastic properties

The Nb$_2$AC (A = S, Sn) phases possess the hexagonal structure which crystallizes in the space group $P6_3/mmc$ (No.194) and has 8 atoms in one unit cell, where blocks of transition metal carbides [NbC] (formed by edge-shared Nb$_6$C octahedra) are sandwiched with A atomic sheets (Fig. 1). The Wyckoff positions of atoms in Nb$_2$AC are − C: 2$a$ [(0, 0, 0) , (0, 0, 1/2)], A: 2$d$ [(1/3, 2/3, 3/4), (2/3, 1/3, 1/4)], and Nb atoms: 4$f$ [(1/3, 2/3, $z_{Nb}$), (2/3, 1/3, $z_{Nb}$+1/2), (2/3, 1/3,−$z_{Nb}$), (1/3, 2/3, −$z_{Nb}$+1/2)] (see [2, 3]).

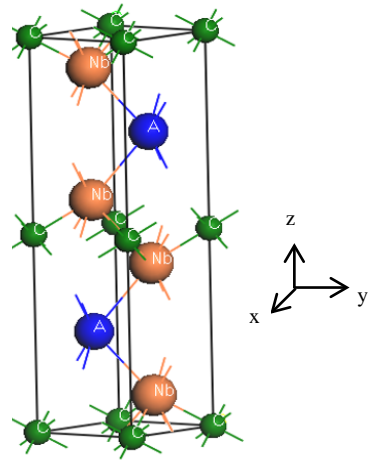

**Fig. 1.** Crystal structure of layered MAX phases: Nb$_2$AC (A = S, Sn).

The calculated fully relaxed equilibrium values of the structural parameters of the two phases are presented in Table 1 together with other theoretical [13-17] and experimental results [5, 6, 19]. The comparison shows that the calculated values are in good agreement with the available experimental results.

**Table 1.** Calculated lattice parameters ($a$ and $c$, in Å), ratio $c/a$, and internal parameters $z_{Nb}$ for MAX phases Nb$_2$SC and Nb$_2$SnC.

| Phases | $a$ | $c$ | $c/a$ | $z_{Nb}$ | Ref |
|--------|-----|-----|-------|----------|-----|
| Nb$_2$SC | 3.290 | 11.670 | 3.550 | 0.0956 | This |
| | 3.294 | 11.553 | 3.510 | 0.0964 | [5] Exp. |
| | 3.294 | 11.783 | 3.577 | 0.0960 | [13] |
| | 3.320 | 11.709 | 3.530 | 0.0952 | [15] |
| Nb$_2$SnC | 3.260 | 13.910 | 4.270 | 0.0820 | This |
| | 3.220 | 13.707 | 4.256 | | [6] Exp. |
| | 3.241 | 13.802 | 4.258 | | [19] Exp. |
| | 3.252 | 13.844 | 4.257 | 0.0830 | [13] |
| | 3.200 | 13.534 | 4.228 | 0.0846 | [14] |
| | 3.277 | 13.903 | 4.242 | 0.0828 | [15] |
| | 3.263 | 13.906 | 4.262 | 0.0821 | [16] |
| | 3.244 | 13.754 | 4.240 | 0.0830 | [17] |

The bulk modulus $B$, shear modulus $G$, Young's modulus $Y$ (all in GPa), and Poisson's ratio $\nu$ at zero pressure are evaluated using calculated elastic constants and presented in Table 2. The ductility of a material can be roughly estimated by the ability of performing shear deformation, such as the value of shear-modulus-to-bulk-modulus ratios. Thus a ductile plastic solid would show low $G/B$ ratio ($< 0.5$); otherwise, the material is brittle. As is evident from Table 2, the calculated $G/B$ ratios for Nb$_2$SC and Nb$_2$SnC phases are equal ($\sim$0.54), indicating that the two compounds are near borderline of ductility. The same can be inferred from an additional argument that the variation in the brittle/ductile behavior follows from the calculated Poisson's ratio. For brittle material the value is small enough, whereas for ductile metallic materials $\nu$ is typically 0.33 [22].

**Table 2.** Calculated $C_{ij}$, (in GPa), $B$ (in GPa), $G$ (in GPa), $Y$ (in GPa), $\nu$, $A$, $A_1$, $k_c/k_a$ for Nb$_2$SC and Nb$_2$SnC.

| Phases | $C_{11}$ | $C_{12}$ | $C_{13}$ | $C_{33}$ | $C_{44}$ | $B$ | $G$ | $Y$ | $\nu$ | A | $A_1$ | $k_c/k_a$ | Ref |
|--------|----------|----------|----------|----------|----------|-----|-----|-----|-------|---|-------|-----------|-----|
| Nb$_2$SC | 320.1 | 100.8 | 152.5 | 327.2 | 125.2 | 196.8 | 107.7 | 273.2 | 0.270 | 1.14 | 1.46 | 0.66 | This |
| | 303.6 | 116.9 | 155.1 | 315.7 | 88.1 | 220.9 | 88.6 | 234.5 | 0.323 | 0.94 | 1.14 | 0.69 | [15] |
| Nb$_2$SnC | 265.3 | 94.6 | 122.6 | 261.7 | 108.2 | 163.4 | 88.9 | 225.8 | 0.270 | 1.26 | 1.53 | 0.82 | This |
| | 315.0 | 99.0 | 141.0 | 309.0 | 124.0 | 210.2[*] | 106.8 | 274.0[*] | 0.282[*] | 1.15[t] | 1.45[t] | 0.78[t] | [14] |
| | 254.8 | 100.8 | 120.0 | 243.0 | 58.9 | 171.1 | 66.8 | 177.3 | 0.327 | 0.77 | 0.91 | 0.94 | [15] |
| | 286.5 | 91.5 | 126.8 | 288.5 | 99.8 | 171.1 | 93.3 | 236.9 | 0.271 | 1.02[t] | 1.24[t] | 0.77[t] | [16] |
| | 340.9 | 105.6 | 169.0 | 320.6 | 183.3 | 209.0 | 125.8 | 314.5 | 0.250 | 1.56[t] | 2.27[t] | 0.72[t] | [17] |
| | 268.0 | 86.0 | 119.0 | 267.0 | 98.0 | 161.0 | 89.0 | 225.4[t] | 0.266[t] | 1.08[t] | 1.32[t] | 0.78[t] | [18] |

[*]Errors corrected using the published data [14].
[t]Estimated using published data of respective authors.

The elastic shear anisotropy of crystals, defined by $A = 2C_{44}/(C_{11}-C_{12})$, can be responsible for the development of microcracks in the material [23]. The factor $A_1 = 4C_{44}/(C_{11}+C_{33}-2C_{13})$ is also used as a measure of the degree of anisotropy for the {100} shear planes between [011] and [010] directions [15]. According to our calculations the value of $A$ increases from 1.14 to 1.26 as S is replaced by Sn. Our estimations demonstrate that the value of $A_1$ for $Nb_2SC$ and $Nb_2SnC$ phases are 1.46 and 1.53, respectively. So we conclude that the considered phases are anisotropic. Yet another anisotropy parameter that can be estimated as the ratio between the uniaxial compression values along the $c$ and $a$ axis for a hexagonal crystal: $k_c/k_a = (C_{11}+C_{12}-2C_{13})/(C_{33}-C_{13})$. Our data show that the compressibility along the $c$-axis is smaller than along the $a$-axis ($k_c/k_a = 0.66$) for Nb2SC (= 0.66) and also for $Nb_2SnC$ (= 0.82), which are in agreement with other calculations [14-17].

### 3.2. Electronic and bonding properties

Figs. 2 (a, b) present the energy bands of $Nb_2SC$ and $Nb_2SnC$ along the high symmetry directions of the Brillouin zone in the energy range from −17 to +5 eV. The band structures of the two nanolaminates are to some extent similar. As observed before [13] they mainly differ in the energy positions of the $s$ and $p$ bands of S or Sn and in the population of the common valence band (VB) due to different electron concentrations, being 40e and 36e per unit cell for $Nb_2SC$ and $Nb_2SnC$, respectively. The valence and conduction bands of $Nb_2SC$ and $Nb_2SnC$ are seen to overlap, thus indicating metallic-like behavior of both the phases.

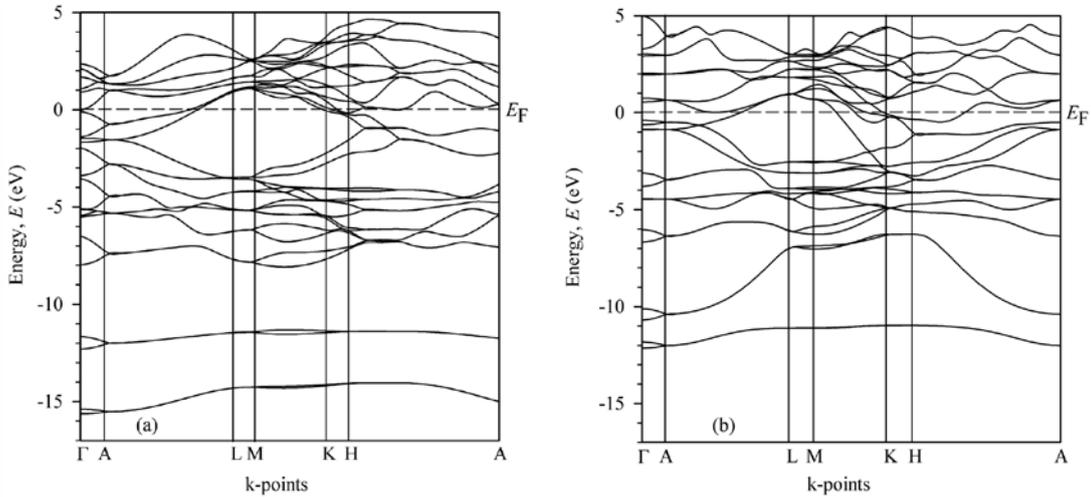

**Fig. 2.** Electronic band structures of (a) $Nb_2SC$ and (b) $Nb_2SnC$.

Figs. 3 (a, b) show the total and partial densities of states for the two compounds $Nb_2SC$ and $Nb_2SnC$ nanolaminates. The density of states at the Fermi level are 3.54 and 3.82 states/eV which predominantly contain contributions from the Nb 4$d$ states of 2.88 and 2.90 states/eV of $Nb_2SC$ and $Nb_2SnC$, respectively.

The VB of $Nb_2SC$ has several energy bands (Fig. 3 a): the quasi-core S 3$s$ bands (from −16 to −13.5 eV) are located below the C 2$s$ bands (−12.5 to −11.8 eV). These are nonbonding quasi-core bands with small energy dispersion. The upper part of the VB (from –8 eV to Fermi level) is contributed to by the C 2*p*, S 3*p*, and Nb 4*d* states. The lower part of the VB is characterized by mixing of the C 2*p*, S 3*p*, and Nb 4*d* states, whereas the near-Fermi region (from –3.6 to Fermi level) is mostly of Nb 4*d* character.

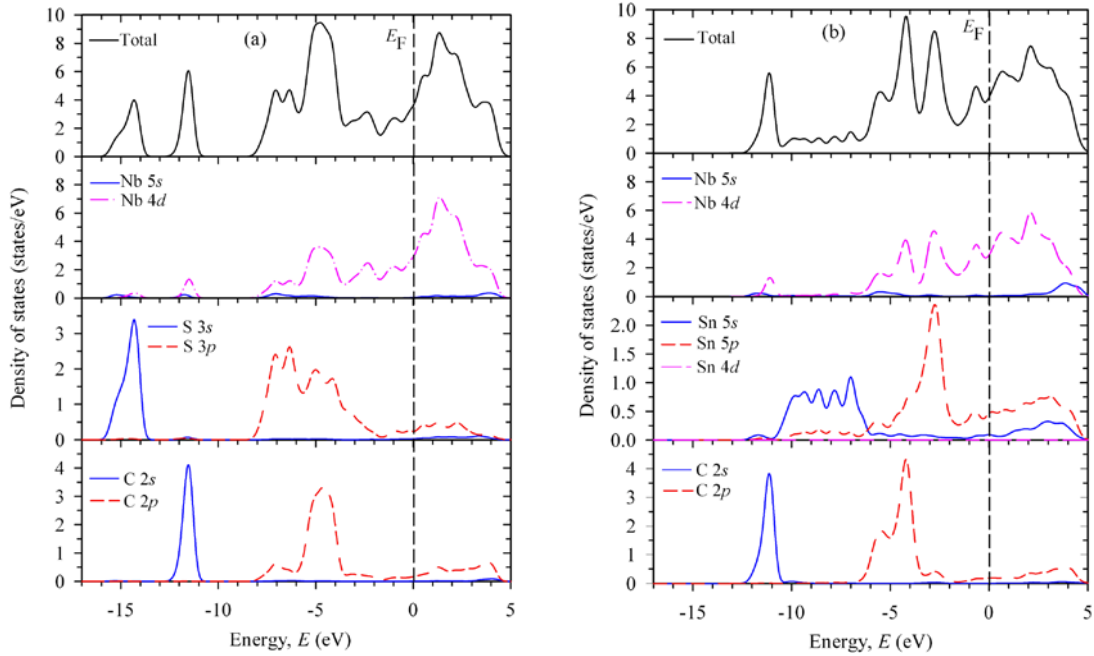

**Fig. 3.** Total and partial DOS of (a) Nb₂SC and (b) Nb₂SnC.

These may now be compared with those for Nb₂SnC (i.e. when S is replaced by Sn) (Fig. 3 b). Here the lower band consisting mainly of C 2*s* states is from –12.3 to –10.5 eV below the Fermi level. These are again quasi-core nonbonding bands with small dispersion. In agreement with the results obtained using different calculational methodology [13] the next two bands, of Sn 5*s* type (from –11 to – 6.5 eV), are of quasi-2D character along the M–K direction, but with a maximum dispersion along the Γ–K and Γ–M directions. This indicates that the Sn 5*s* orbitals are involved in Sn–Sn bonding (in the plane of tin networks) of atoms from neighboring Nb networks [13]. The top of the VB  (from – 6.8 eV to Fermi level) is formed by several hybrid bands mainly made up of the Nb 4*d*, C 2*p*, and Sn 5*p* orbitals. It may be mentioned here that the VB part for Nb₂SC is 1.2 eV wider than that for Nb₂SnC as a result of filling of Nb 4*d* bands due to increase in electron concentration. Further, the Fermi level for both the phases falls in the region of nonbonding Nb 4*d* states.

The top of the VB (from –8 eV to Fermi level) is formed by hybridizing bands made up of the Nb 4*d*, C 2*p*, and S 3*p* orbitals in case of Nb₂SC. But for Nb₂SnC the top of the VB  (from –6.8 eV to Fermi level) is formed by hybridizing bands made up of the Nb 4*d*, C 2*p*, and Sn 5*p* orbitals. These Nb 4*d* –S 3*p* (Nb 4*d* –Sn 5*p* for Nb₂SnC) hybrids are closer to Fermi level than the Nb 4*d* – C 2*p* ones suggesting that Nb–C bonds are stronger than Nb–S(Sn) bonds. These results are in agreement with the bond lengths obtained from an analysis of Mulliken bond population. The bond lengths in Å are: Nb–C (2.2061, 2.2006), Nb–S (2.6207), Nb–Sn (3.0008); Nb–Nb (2.9330, 2.9568), S–C (3.4835), Sn–C (3.9541).

### 3.3. Thermodynamic properties

We investigated the thermodynamic properties of $Nb_2SC$ and $Nb_2SnC$ by using the quasi-harmonic Debye model in a manner as described elsewhere [32, 33]. Here we computed the bulk modulus, Debye temperature specific heats, and volume thermal expansion coefficient at different temperatures and pressures. For this we utilized $E$-$V$ data obtained from the third-order Birch-Murnaghan equation of state [34] using zero temperature and zero pressure equilibrium values of $E_0$, $V_0$, $B_0$, based on DFT method.

Fig. 4 (a) presents the temperature variation of isothermal bulk modulus of $Nb_2SC$ and $Nb_2SnC$, the *inset* of which shows bulk modulus $B$ data as a function of pressure. Our calculations show that both $B$ values (larger for $Nb_2SC$) are nearly flat for $T < 100K$. For $T > 100K$, $B$ for $Nb_2SC$ phase decreases at a slightly faster rate than that the other phase. The *inset* shows the pressure variation of room temperature bulk modulus. It is found that $B$ increases with pressure at a given temperature and decreases with temperature at a given pressure, which is consistent with the trend of volume.

The temperature dependence of Debye temperature $\Theta_D$ of $Nb_2SC$ and $Nb_2SnC$ at zero pressure is displayed in Fig. 4 (b). We note that $\Theta_D$, smaller for $Nb_2SnC$, decrease non-linearly with temperature for both the phases. The pressure dependent Debye temperature (presented in the *inset* at $T = 300K$) shows a non-linear increase. The variation of $\Theta_D$ with pressure and temperature reveals that the thermal vibration frequency of atoms in the nanolaminates changes with pressure and temperature.

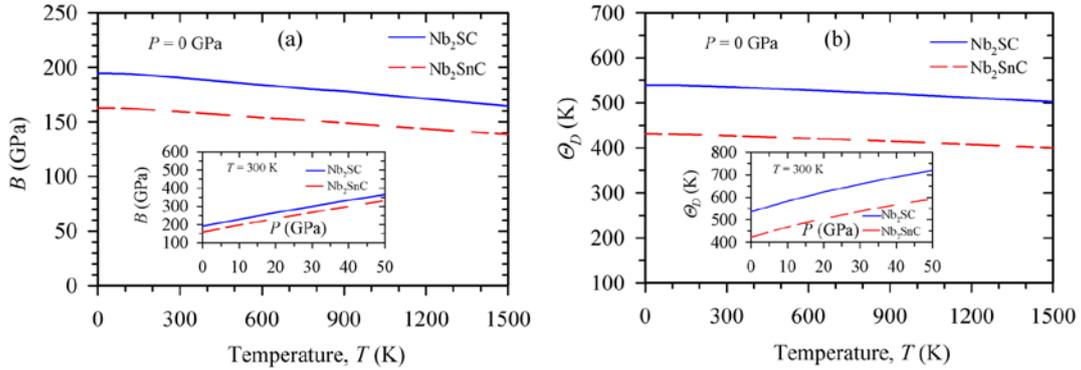

**Fig. 4.** Temperature dependence of (a) Bulk modulus and (b) Debye temperature of $Nb_2SC$ and $Nb_2SnC$. *Inset* shows pressure variation.

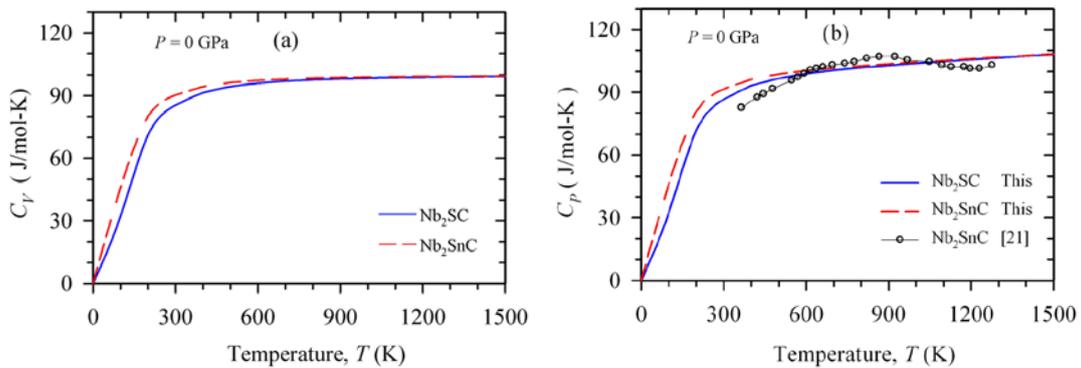

**Fig. 5.** Temperature dependence of (a) specific heat at constant volume, and (b) specific heat at constant pressure of $Nb_2SC$ and $Nb_2SnC$.

The constant-volume and constant-pressure specific heat capacities $C_V$, $C_P$ of Nb$_2$SC and Nb$_2$SnC as a function of temperature are shown in Figs. 5 (a, b). The heat capacities increase with increasing temperature, because phonon thermal softening occurs when the temperature increases. It may be noted that the heat capacity anomaly close to $T_c$-value (5 and 7.8 K, for the two superconductors) is of the order of mJ that it has no effect on the analysis being made here. The difference between $C_P$ and $C_V$ in the normal state for the phases is given by $C_P – C_V = \alpha_V^2(T)\ BTV$ ($\alpha_V$ = volume thermal expansion coefficient, VTEC), which is due to the thermal expansion caused by anharmonicity effects. In the low temperature limit, the specific heat exhibits the Debye $T^3$ power-law behavior and at high temperature ($T > 300$K) the anaharmonic effect on heat capacity is suppressed, and $C_V$ approaches the classical asymptotic limit of $C_V = 3nNk_B = 99.7$ J/mol-K. These results show the fact that the interactions between ions in the nanolaminates have great effect on heat capacities especially at low temperatures.

The thermal properties of Nb$_2$SnC including specific heat at constant pressure have been reported by Barsoum *et al.* [21]. We show the measured $C_P$ values as a function of temperature in Fig. 5 (b). The agreement is found to be satisfactory. Now the calculated data on DOS, $N(E_F)$ enable us to estimate the Sommerfeld constant $\gamma$ (giving the electronic contribution to specific heat) within the free electron model using $\gamma = (\pi^2/3)\ N(E_F)$; the resulting values are 5.45 mJ K$^2$ mol$^{-1}$ and $\gamma = 5.88$ (5.7, see [24]) mJ K$^2$ mol$^{-1}$ for Nb$_2$SC and Nb$_2$SnC, respectively. We can also estimate the electron-phonon coupling constant ($\lambda$) using McMillan's relation [35]:

$$\lambda = \frac{1.04 + \mu^* \ln\left(\frac{\Theta_D}{1.45\,T_c}\right)}{(1 - 0.62\,\mu^*) \ln\left(\frac{\Theta_D}{1.45\,T_c}\right) - 1.04} \tag{1}$$

where $\mu^*$ is a Coulomb repulsion constant (typical value, $\mu^* = 0.10$). Taking the measured $T_c$ values, and the calculated Debye temperatures, we find $\lambda \sim 0.49$, and $\sim 0.59$, for Nb$_2$SC and Nb$_2$SnC, respectively. The values imply that both the phases are moderately coupled superconductors.

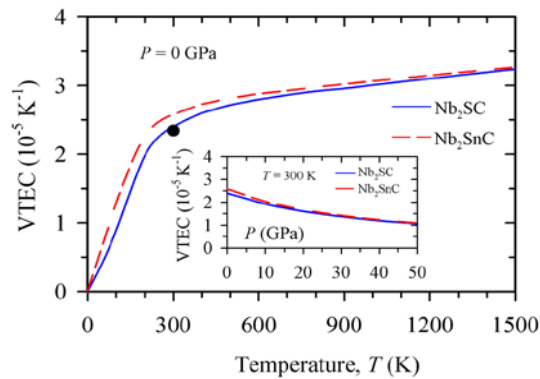

**Fig. 6.** **T**emperature dependent volume expansion co-efficient of Nb$_2$SC and Nb$_2$SnC. *Inset* shows pressure variation. The measured value for Nb$_2$SnC [26] is shown as a filled circle.

Fig. 6 shows the volume thermal expansion coefficient (VTEC), $\alpha_V$ as a function of temperature and pressure (*inset*). The thermal expansion coefficient increases rapidly especially at temperature below 300K, whereas it gradually tends to a slow increase at higher temperatures. On the other hand at a constant temperature, the expansion coefficient decreases strongly with pressure. It is well-known that the thermal expansion coefficient is inversely related to the bulk modulus of a material. Further the coefficient of the $M_2AC$ phases scale with those of the corresponding stoichiometric transition metal binary carbides or MC [21]. The coefficients are thus quite low for such readily machinable solids; the measured lowest value ($7.8 \times 10^{-6}$ $K^{-1}$) is that of $Nb_2SnC$, and the highest ($10 \times 10^{-6}$ $K^{-1}$) belongs to $Ti_2SnC$ [26]. Taking $\alpha_V = 3\times$ (linear thermal expansion coefficient), our calculated value of $\alpha_V = 2.55 \times 10^{-5} K^{-1}$ for $Nb_2SnC$ around room temperature yields $\alpha = 8.5 \times 10^{-6} K^{-1}$ which is reasonable.

### 3.4. Optical properties

The dielectric function $\varepsilon(\omega) = \varepsilon_1(\omega) + i\varepsilon_2(\omega)$ fully describes the optical properties of any homogeneous medium at all photon energies. The imaginary part $\varepsilon_2(\omega)$ is obtained from the momentum matrix elements between the occupied and the unoccupied electronic states and calculated directly using [36]:

$$\varepsilon_2(\omega) = \frac{2e^2\pi}{\Omega\varepsilon_0} \sum_{k,v,c} \left| \psi_k^c \left| \boldsymbol{u}\boldsymbol{r} \right| \psi_k^v \right|^2 \delta\left( E_k^c - E_k^v - E \right) \qquad (2)$$

where $\boldsymbol{u}$ is the vector defining the polarization of the incident electric field, $\omega$ is the light frequency, $e$ is the electronic charge and $\psi_k^c$ and $\psi_k^v$ are the conduction and valence band wave functions at $k$, respectively. In the very low energy infrared spectra the effect of Drude term is nonnegligible and the correction for it is made [36]. The real part is derived from the imaginary part $\varepsilon_2(\omega)$ by the Kramers-Kronig transform. All other optical constants, such as refractive index, absorption spectrum, loss-function, reflectivity and conductivity (real part) are those given by Eqs. 49 to 54 in ref. [36].

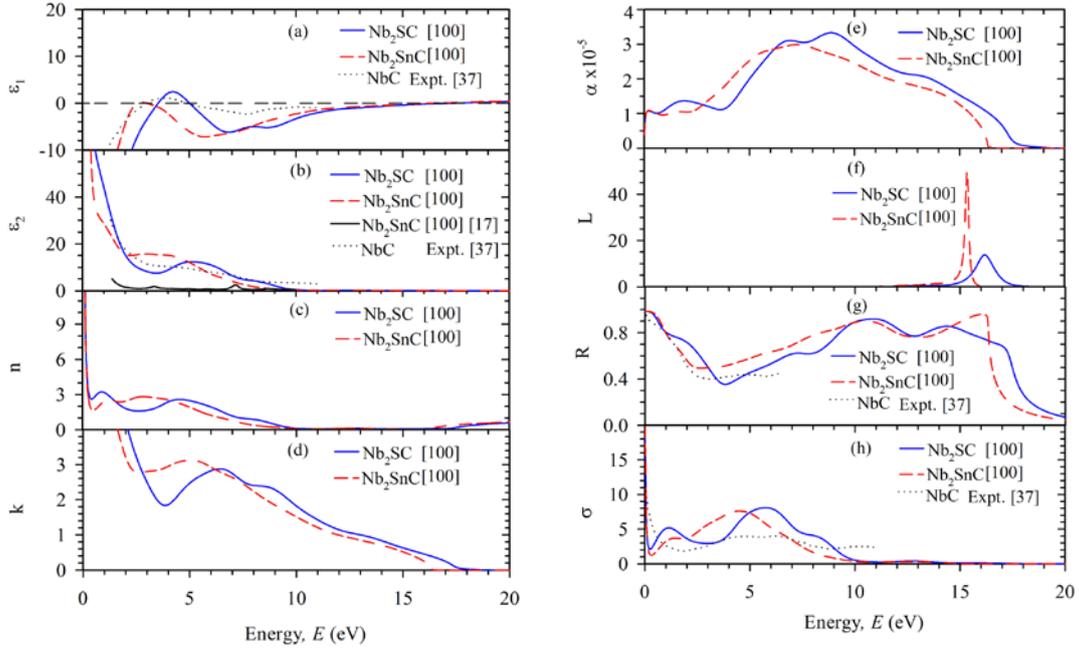

**Fig. 7.** (a) Real part of dielectric function, (b) imaginary part of dielectric function, (c) real part of refractive index, (d) extinction coefficient, (e) absorption, (f) loss function, (g) reflectivity, and (h) real part of conductivity of $Nb_2SC$ and $Nb_2SnC$ in [100] direction.

Fig. 7 shows the optical functions of Nb$_2$SC and Nb$_2$SnC calculated for photon energies up to 20 eV for polarization vectors [100] and [001] (only spectra for [100] shown), along with a lone theoretical spectrum of dielectric function of Nb$_2$SnC [17]. We have used a 0.5 eV Gaussian smearing for all calculations. This smears out the Fermi level, so that *k*-points will be more effective on the Fermi surface. A Drude term with unscreened plasma frequency 7 eV and damping 0.05 eV has been used.

The real and imaginary parts of dielectric function of the two nanolaminates are shown Fig. 7 (a) and 7 (b) along with the curve from Kanoun *et al.* [17] which is for a limited energy range and also that of Nb$_2$C$_{0.98}$ [37]. Nb$_2$SC and Nb$_2$SnC exhibit metallic characteristics in the energy ranges for which $\varepsilon_1(\omega) < 0$. As can be seen the result of Nb$_2$SnC is somewhat different as regards the energy ranges for negativity of $\varepsilon_1(\omega)$. For the real part $\varepsilon_1(\omega)$ of the dielectric function, the peaks for Nb$_2$SC and Nb$_2$SnC appear at around 4.25 eV and 2.8 eV, respectively. In the energy range of 1.8 – 6 eV, $\varepsilon_2$ curve due to Kanoun *et al.* [17] has roughly similar characteristics but the magnitude is quite different from our data. The refractive index and extinction coefficient are displayed in Figs. 7 (c, d).

The absorption spectra shown in Fig. 7 (e) reveal the metallic nature of the nanolaminates since the spectra starts from 0 eV. Nb$_2$SnC has three peaks at 1.9, 6.7, 9.0 eV, besides having a shoulder at ~13.5 eV, whereas Nb$_2$SnC has a broad main peak at ~ 7 eV. The function $L(\omega)$, shown in Fig. 7 (f), describes the energy loss of a fast electron traversing in the material. Its peak is defined as the bulk plasma frequency $\omega_P$, which occurs at $\varepsilon_2 < 1$ and $\varepsilon_1 = 0$. In the energy-loss spectrum, we see that the effective plasma frequency $\omega_P$ of the two phases are equal to 16.2 and 15.4 eV. When the incident photon frequency is higher than $\omega_P$, the material becomes transparent.

The reflectivity spectra as a function of photon energy are shown in Fig. 7 (g) in comparison with measured spectra of NbC$_{0.98}$. It is found that the reflectivity of both the compounds, having nearly similar characteristics, starts with a value of ~ 0.9-0.98, decreases and then rises again to reach maximum value of ~0.8 to 0.9 between 10 – 16.5 eV. Thus both the phases, with roughly similar reflectivity spectra, show promise as good coating materials between 10-16.5 eV regions. Since the material has no band gap as evident from band structure, the photoconductivity starts with zero photon energy as shown in Fig. 7 (h). The measured spectra of NbC [37] are shown in the figure for comparison. The photoconductivity and hence electrical conductivity of the materials increases as a result of absorbing photons.

## 4. Conclusion

First-principles calculations based on DFT have been used to study the structural, elastic, electronic, thermodynamic, and optical properties of the two MAX phases Nb$_2$SC and Nb$_2$SnC. The calculated elastic constants and various elastic anisotropies of Nb$_2$SC and Nb$_2$SnC are discussed and compared with available calculations. The two nanolaminates are found to be quasi-ductile in nature.

Band structure and total densities of states analysis suggest that both materials exhibit metallic conductivity which increases as A is changed from S to Sn in Nb$_2$AC. The bonding is achieved through hybridizing bands made up of the Nb 4*d*, C 2*p*, and S 3*p*/Sn 5*p* orbitals for Nb$_2$SC and Nb$_2$SnC.

The finite-temperature ($\leqslant$ 1500K) and finite-pressure ($\leqslant$ 50GPa) bulk modulus, specific heats, thermal expansion coefficient, and Debye temperature are all obtained and the results analyzed. The variation of $\Theta_D$ with temperature and pressure reveals the changeable vibration frequency of the particles in the two nanolaminates. The heat capacities increase with increasing temperature, which shows that phonon thermal softening occurs when the temperature increases. The $C_P$ and the thermal expansion coefficient are in agreement with the measured values available only for $Nb_2SnC$. The estimation yields the electron-phonon coupling strengths $\lambda \sim 0.49$ ($Nb_2SC$), and 0.59 ($Nb_2SnC$), which indicate that both the phases are moderately coupled superconductors.

From an analysis of optical functions, it is found that $Nb_2SC$ is a better dielectric material. Further the reflectivity is seen to be high in visible-ultraviolet regions between 9-15 eV for both $Nb_2SC$ and $Nb_2SnC$, showing promise as good coating materials.